\documentclass[aps,preprint,nofootinbib,superscriptaddress]{revtex4}

\usepackage{amssymb}

\usepackage{epsfig}
\usepackage[bookmarksnumbered,bookmarksopen,colorlinks,citecolor=blue,linkcolor=blue]{hyperref}

\begin{document}

\title{ Systematic study of $^{16}$O-induced fusions with the improved quantum molecular dynamics model }

\author{Ning Wang}
\email{wangning@gxnu.edu.cn}\affiliation{ Department of Physics,
Guangxi Normal University, Guilin 541004, People's Republic of
China }

\author{Kai Zhao}
\affiliation{China Institute of Atomic Energy, Beijing 102413, People's Republic of
China}

\author{Zhuxia Li}
\affiliation{China Institute of Atomic Energy, Beijing 102413, People's Republic of
China}

\begin{abstract}

The heavy-ion fusion reactions with $^{16}$O bombarding on $^{62}$Ni, $^{65}$Cu, $^{74}$Ge, $^{148}$Nd, $^{180}$Hf, $^{186}$W, $^{208}$Pb, $^{238}$U are systematically investigated with the improved quantum molecular dynamics
(ImQMD) model. The fusion cross sections at energies near and above the Coulomb barriers can be reasonably well reproduced by using this semi-classical microscopic transport model with the parameter sets SkP* and IQ3a. The dynamical nucleus-nucleus potentials and the influence of Fermi constraint on the fusion process are also studied simultaneously. In addition to the mean field, the Fermi constraint also plays a key role for the reliable description of fusion process and for improving the stability of fragments in heavy-ion collisions.

\end{abstract}
\maketitle

\begin{center}
\textbf{I. INTRODUCTION}
\end{center}

The heavy-ion fusion reactions are of significant importance not only for the study of the nuclear structure and test of the models, but also for the synthesis of new super-heavy elements. It is usually thought that the fusion process is a process of tunneling through the Coulomb barrier from the point of view of quantum mechanics. Under the parabolic approximation for the potential barrier, the fusion cross sections for the reactions between two light nuclei can be described by an analytical expression, i.e. the Wong's formula \cite{Wong73}. For the fusion reactions with intermediate and heavy nuclei, the coupling between the relative motion and other degrees of freedom becomes important and the fusion excitation function can be reasonably well reproduced by the coupled-channel calculation program CCFULL \cite{Hag99} together with an empirical nuclear potential. In addition, it is found that the neutron-rich effect, shell effect and nucleon transfer effect can also influence the fusion cross sections at sub-barrier energies. The systematic study of the fusion excitation functions with the Skyrme energy-density functional approach indicates that the neutron-rich effect (due to the transfer of neutrons and the formation of neutron-rich neck) can significantly suppress the fusion barrier and thus cause the enhancement of fusion cross sections at sub-barrier energies, whereas the strong shell effect can suppress the lowering barrier effect \cite{liumin06,Wang09}. In Refs. \cite{Brog83,Zhang10,Zag03,Sarg12}, the authors claimed that the nucleon transfer with positive Q-values can cause the enhancement of fusion cross sections at sub-barrier energies. To explore the influence of dynamical effect on the fusion process, the microscopic dynamics models, such as the time-dependent Hartree-Fock (TDHF) model \cite{Umar06,Umar12} and the improved quantum molecular dynamics (ImQMD) model \cite{ImQMD2002,ImQMD2004,ImQMD2010} have been developed. It is found that the dynamical effect such as the energy-dependence of the potential barrier plays a key role for the fusion process from these microscopic dynamics simulations. From the point of view of the semi-classical ImQMD model based on event-by-event simulations, the "sub-barrier" fusion is a process that the rare projectile nuclei surmount rather than tunnel through the suppressed potential barrier \cite{ImQMD2014}. These different explanations for the fusion cross sections indicate that the  mechanism of heavy-ion fusion reactions is still not very clear and more fusion reactions should be further investigated.

The ImQMD model is a semi-classical microscopic dynamics transport model and is successfully applied on heavy-ion fusion reactions between stable nuclei \cite{ImQMD2004,ImQMD2012} and fusion reactions induced by neutron-rich nuclei \cite{ImQMD2014}. In this work, we would like to further test the ImQMD model for the description of fusion process with more reactions from intermediate system to heavy system. The fusion excitation functions for eight fusion reactions induced by $^{16}$O will be systematically studied with this model and the dynamical nucleus-nucleus potential will be compared. In the earlier works \cite{ImQMD2012, ImQMD2014, ImQMD13}, the influences of the mean field and the initialization of nuclei in the ImQMD model were mainly investigated. In the standard quantum molecular dynamics (QMD) model, the anti-symmetrization of wave function is neglected and thus the fermionic nature of nuclear system can not be reasonably represented. In the ImQMD model, the Fermi constraint \cite{constrain,ImQMD2010}, which is an effective method to describe the fermionic nature of the $N$-body system and to improve the stability of an individual nucleus, is adopted. In addition to the mean field, the Fermi constraint should also be important for the reliable description of the fusion process. It is therefore necessary to investigate the influence of the Fermi constraint on the fusion cross sections and reaction yields.

The structure of this paper is as follows: In sec. II, the framework of the ImQMD model will be briefly introduced. In sec. III, the fusion cross sections of eight fusion reactions induced by $^{16}$O and the dynamical nucleus-nucleus potentials will be presented. The influence of the Fermi constraint on the fusion cross section and charge distribution will also be investigated. Finally a brief summary is given in Sec. IV.

\begin{center}
\noindent{\bf {II. IMPROVED QUANTUM MOLECULAR DYNAMICS MODEL }}\\
\end{center}

In the improved quantum molecular dynamics model (with the version ImQMD2.1, the same version as that used in Ref.\cite{ImQMD2014}), each
nucleon is represented by a coherent state of a Gaussian wave
packet. The density distribution function $\rho$ of a system reads
\begin{equation} \label{1}
\rho(\mathbf{r})=\sum_i{\frac{1}{(2\pi \sigma_r^2)^{3/2}}\exp
\left [-\frac{(\mathbf{r}-\mathbf{r}_i)^2}{2\sigma_r^2} \right ]},
\end{equation}
where $\sigma_r$ represents the spatial spread of the wave packet.
The propagation of nucleons is governed by the self-consistently generated mean field,
\begin{equation} \label{2}
\mathbf{\dot{r}}_i=\frac{\partial H}{\partial \mathbf{p}_i}, \; \;
\mathbf{\dot{p}}_i=-\frac{\partial H}{\partial \mathbf{r}_i},
\end{equation}
where $r_i$ and $p_i$ are the center of the $i$-th wave packet in
the coordinate and momentum space, respectively.  The Hamiltonian
$H$ consists of the kinetic energy
$T=\sum\limits_{i}\frac{\mathbf{p}_{i}^{2}}{2m}$ and the effective
interaction potential energy $U$:
\begin{equation} \label{3}
H=T+U.
\end{equation}
The effective interaction potential energy is written as the sum
of the nuclear interaction potential energy $U_{\rm
loc}=\int{V_{\rm loc}(\textbf{r})d\textbf{r}}$ and the Coulomb
interaction potential energy $U_{\rm Coul}$ which includes the
contribution of the direct and exchange terms,
\begin{equation}
U=U_{\rm loc}+U_{\rm Coul}.
\end{equation}
Where $V_{\rm loc}(r)$ is the potential energy density that is
obtained from the effective Skyrme interaction without the spin-orbit term:
\begin{equation}
V_{\rm
loc}=\frac{\alpha}{2}\frac{\rho^2}{\rho_0}+\frac{\beta}{\gamma+1}\frac{\rho^{\gamma+1}}{\rho_0^{\gamma}}+\frac{g_{\rm
sur}}{2\rho_0}(\nabla\rho)^2
+g_{\tau}\frac{\rho^{\eta+1}}{\rho_0^{\eta}}+\frac{C_s}{2\rho_0}[\rho^2-k_s(\nabla\rho)^2]\delta^2
\end{equation}
where $\delta=(\rho_n -\rho_p)/(\rho_n +\rho_p)$ is the isospin
asymmetry. In Table I we list the two sets of model parameters adopted in the calculations. The corresponding value of the incompressibility coefficient of nuclear matter is $K_{\infty} = 195$ and 225 MeV for SkP* and IQ3a, respectively.

\begin{table}
 \caption{ Model parameters adopted in the ImQMD calculations.}
\begin{tabular}{lccccccccccc}
\hline Parameter & $\alpha $ & $\beta $ & $\gamma $ &$%
g_{\rm sur}$ & $ g_{\tau }$ & $\eta $ & $C_{s}$ & $\kappa _{s}$ &
$\rho
_{0}$ & ~~$\sigma_0$~~ & ~~$\sigma_1$~~ \\
 & (MeV) & (MeV) &  & (MeVfm$^{2}$) & (MeV) &  & (MeV) & (fm$^{2}$) &
 (fm$^{-3}$) & (fm) & (fm) \\ \hline
 SkP* & $-356$ & 303 & 7/6 & 19.5 & 13 & 2/3 &  35  & 0.65  & 0.162 & 0.94 & 0.018\\
IQ3a & $-207$ & 138 & 7/6 & 16.5 & 14 & 5/3 &  34  & 0.4  & 0.165 & 0.94 & 0.02\\
  \hline
\end{tabular}
\end{table}

To describe the fermionic nature of the $N$-body system and to improve the stability of an individual nucleus, the Fermi constraint is simultaneously adopted. According to the Pauli principle, the phase space occupation number $\bar f_{i}$ should be smaller than or equal to one for the $i$-th particle. In the standard QMD simulations, the value of $\bar f_{i}$ could be larger than one in some cases due to the neglecting of the anti-symmetrization of wave function. In the Fermi constraint, the phase space occupation numbers are checked during the propagation of nucleons. If $\bar f_{i}>1$, the momentum of the particle $i$ are randomly changed by a series of two-body elastic scattering between this particle and its neighboring particles which guarantee that the total momentum and total kinetic energy are conserved in the procedures. The Pauli blocking condition and the total energy of the system at the next time step are simultaneously checked. The initialization of the ImQMD simulations is as the same as that adopted in Ref. \cite{ImQMD2014} and the collision term is not involved in the calculations.

\begin{center}
\textbf{III. RESULTS AND DISCUSSIONS}
\end{center}

In this section, the fusion excitation functions of eight fusion reactions induced by $^{16}$O from the ImQMD simulations will be firstly presented. Then,
the dynamical nucleus-nucleus potentials will be calculated and compared with the empirical barrier distribution functions. Finally, the influence of the Fermi constraint on the fusion cross section and charge distribution will be investigated.

\begin{center}
\textbf{A. Fusion excitation function}
\end{center}

\begin{figure}
\includegraphics[angle=0,width=0.9\textwidth]{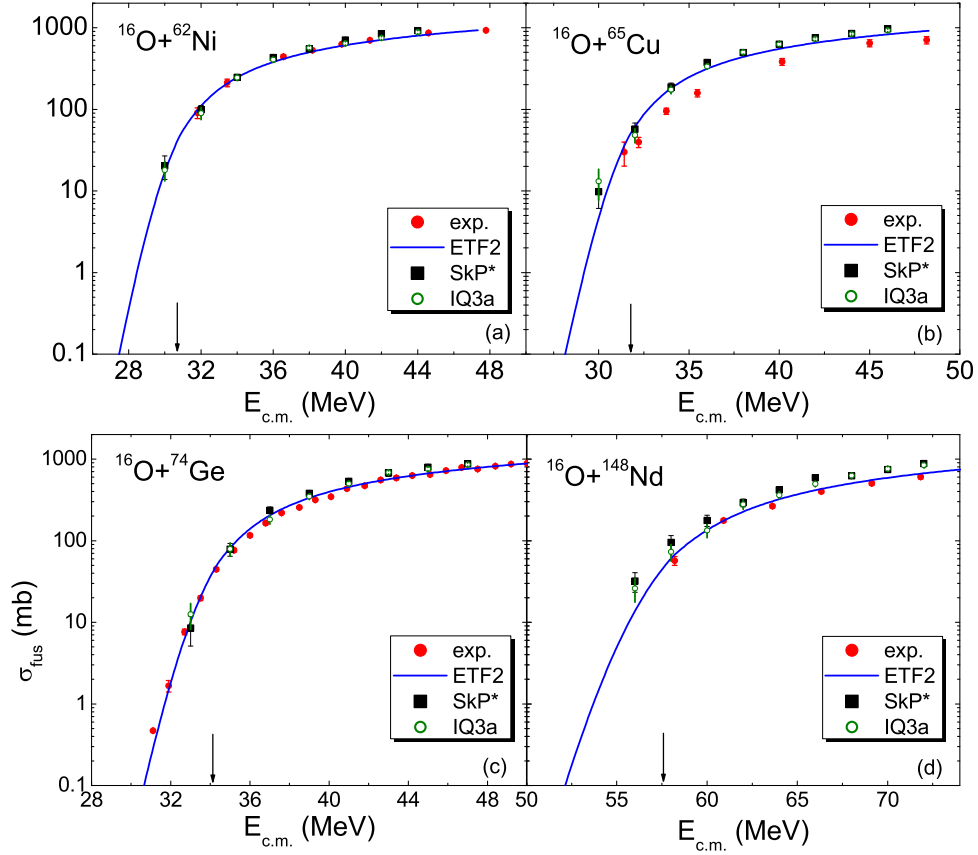}
\caption{(Color online) Fusion excitation functions of $^{16}$O+$^{62}$Ni,$^{65}$Cu,$^{74}$Ge,$^{148}$Nd. The solid circles denote the experimental data taken from \cite{Kee98,Cham92,OGe,ONd}, respectively. The blue curves denote the results with an empirical barrier distribution in which the fusion barrier is obtained by using the Skyrme energy-density functional together with the extended Thomas-Fermi (ETF2) approximation \cite{liumin06,Wang09}. The solid squares and open circles denote the results of ImQMD with the parameter set SkP* and IQ3a, respectively. The statistical errors in the ImQMD calculations are given by the error bars. }
\end{figure}

\begin{figure}
\includegraphics[angle=0,width=0.9\textwidth]{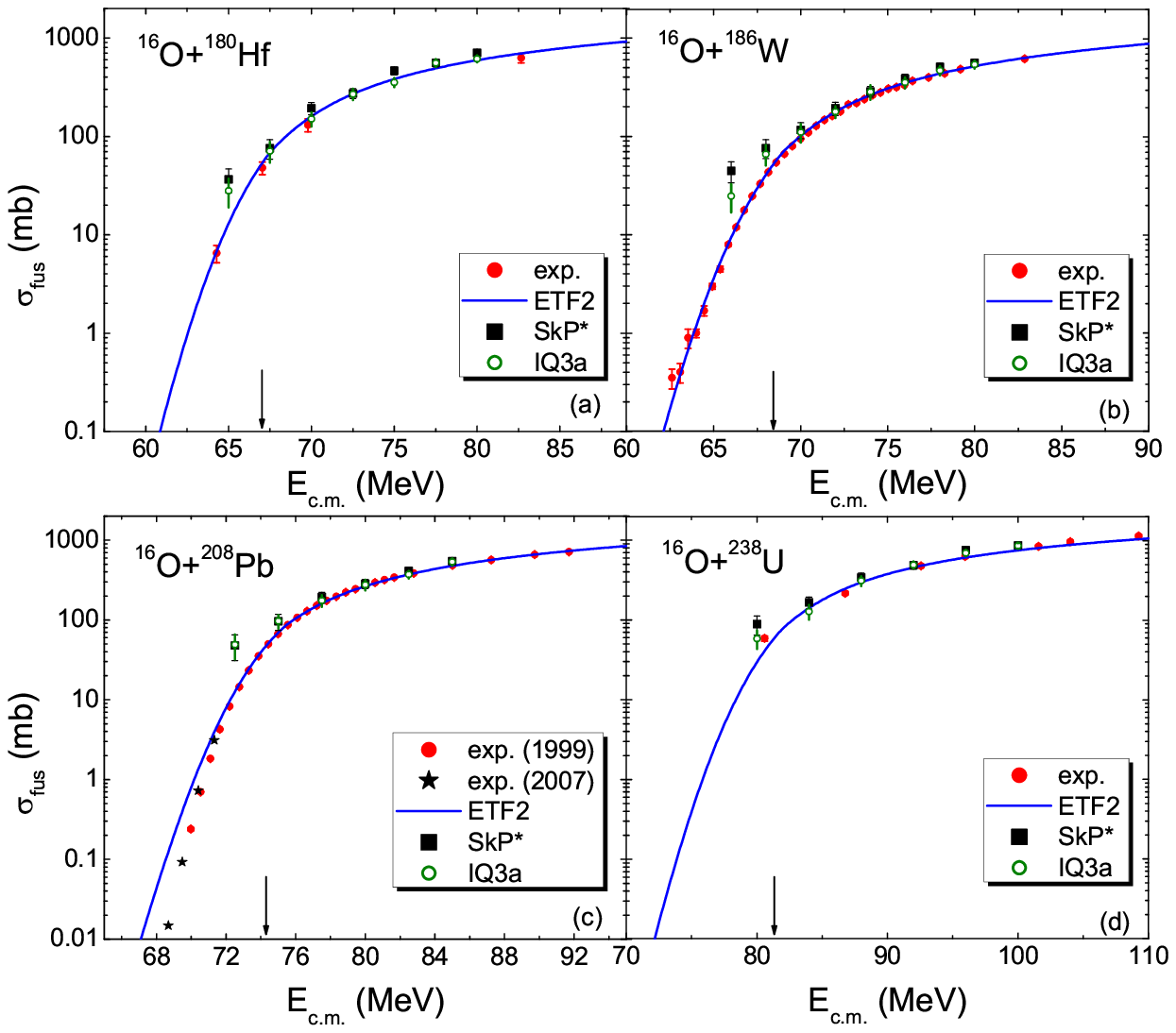}
\caption{(Color online) The same as Fig.1, but for reactions $^{16}$O+$^{180}$Hf,$^{186}$W,$^{208}$Pb,$^{238}$U. The experimental data are taken from Refs. \cite{OHf,OSm,OPb,OPb2,OU}. Here, the initial distance between the reaction partners at z direction (beam direction) is taken to be $d_0 = 40$ fm.}
\end{figure}

Through creating certain bombarding events (about 100 to 200) at each incident energy $E_{\rm c.m.}$ and each impact parameter $b$, and counting the number of fusion events, we obtain the fusion probability $g_{\rm fus}(E_{\rm c.m.},b)$ for a certain fusion reaction. The corresponding fusion excitation function can be calculated with
\begin{equation}
\sigma _{\rm fus}(E_{\rm c.m.})=2\pi \int b \, g_{\rm fus} \, db
\simeq 2\pi \sum b \, g_{\rm fus} \, \Delta b.
\end{equation}
Where, we set $\Delta b=1$ fm. In the calculation of the fusion probability, the event is counted as a fusion (capture) event if the center-to-center distance between the two nuclei is smaller than the nuclear radius of the compound nuclei (which is much smaller than the fusion radius), and the number of bombarding events increases with the decreasing of the incident energies. Here, the quasi-fission probability is neglected for the considered reaction systems.

In Ref. \cite{ImQMD2014}, the fusion reactions $^{16}$O+$^{46}$Ti, $^{16}$O+$^{56}$Fe, $^{16}$O+$^{92}$Zr, $^{16}$O+$^{154}$Sm were investigated with the same model. The fusion cross sections at energies around the Coulomb barrier can be well reproduced. To perform a systematic investigation, we study some other fusion reactions induced by $^{16}$O in this work. Figure 1 and 2 show the comparison of the calculated results and the experimental data for the fusion reactions  $^{16}$O+$^{62}$Ni, $^{16}$O+$^{65}$Cu, $^{16}$O+$^{74}$Ge, $^{16}$O+$^{148}$Nd, $^{16}$O+$^{180}$Hf, $^{16}$O+$^{186}$W, $^{16}$O+$^{208}$Pb, $^{16}$O+$^{238}$U. The solid circles denote the experimental data. The blue curves denote the results with an empirical barrier distribution in which the fusion barrier is obtained by using the Skyrme energy-density functional together with the extended Thomas-Fermi (ETF2) approximation \cite{liumin06,Wang09}. The arrows denote the corresponding most probable barrier height according to the barrier distribution function in the ETF2 approach. One sees that the experimental data can be remarkable well reproduced with the ETF2 approach. With the microscopic dynamics model, the experimental data at energies near and above the Coulomb barrier can be reasonably well reproduced, whereas the data at sub-barrier energies are over-predicted for some reactions with heavy targets, which is probably due to the slightly over-predicted surface diffuseness of nuclei and the neglecting of the shell effects in the self-consistently dynamical evolutions. The results with the parameter set IQ3a are slightly better than those with SkP* due to the relatively smaller value for the surface coefficient $g_{\rm sur}$.  Although the present version of the ImQMD model can not reasonably well describe the sub-barrier fusion of reactions with some heavy spherical targets due to the neglecting of the shell effects, it is still helpful to investigate the dynamical mechanism of the reactions based on this self-consistent microscopic dynamics model, especially for the reactions at intermediate energies and the fusion reactions between two heavy nuclei in which the capture pocket could disappear, since there are not any adjustable model parameters and/or additional assumptions for the reaction process in the whole simulations.

For the heavy-ion fusion reactions, the nuclear surface diffuseness and the dynamical deformation of the reactions partners significantly affect the fusion cross sections at sub-barrier energies. In the present version of ImQMD model, the surface diffuseness of heavy nuclei is slightly over-predicted due to the
approximate treatment of the Fermionic properties of nuclear system, which causes the over-predicted fusion cross sections at sub-barrier energies for the reactions with heavy target nuclei. In addition, for the fusion reactions with doubly-magic nuclei such as $^{208}$Pb, the fusion cross sections at sub-barrier energies are significantly over-predicted by the ImQMD calculations due to the neglecting of the shell effects. The strong shell effect of nuclei can inhibit the dynamical deformation and nucleon transfer, and therefore inhibit the lowering barrier effect. For some neutron-rich fusion systems such as $^{40}$Ca+$^{96}$Zr and $^{132}$Sn+$^{40}$Ca, the results of the ImQMD model are relatively better \cite{ImQMD2014}, which could be due to that the neutron-rich effect is more evident than the shell effect in these reactions.

In addition, it is known that the deformation and orientation effects are important for sub-barrier fusion induced by deformed nuclei \cite{Hag99,Fer89,Kumar11} such as $^{180}$Hf, $^{186}$W, and $^{238}$U. In the semi-classical ImQMD model, the ground state deformations of nuclei can not be self-consistently described due to the neglecting of the shell effect. For the heavy-ion fusion reactions induced by deformed nuclei at energies above the Coulomb barrier, we find that the measured fusion (capture) cross sections can be reasonably well reproduced with the semi-classical ImQMD model, even the deformations and orientation effects are not taken into account additionally. It could be due to that the deformed nucleus with various orientations in the realistic fusion reactions may be approximately described with an equivalent spherical nucleus for the reactions at energies above the barrier. For sub-barrier fusion, the dynamical deformations, in addition to the ground state static deformations of nuclei, also plays a role to the fusion barrier and fusion cross sections. The dynamical deformations of nuclei can be self-consistently described with the ImQMD model and the influence of the dynamical deformations on the nucleus-nucleus potential will be further discussed in the next sub-section.

\begin{center}
\textbf{B. Dynamical nucleus-nucleus potential}
\end{center}

\begin{figure}
\includegraphics[angle=0,width=0.7 \textwidth]{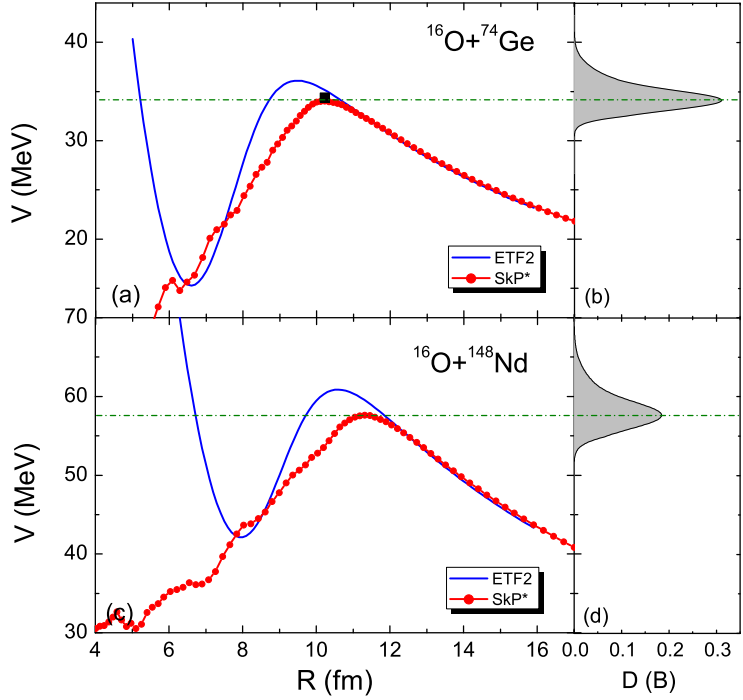}
\caption{(Color online) (a) and (c): Nucleus-nucleus potential for $^{16}$O+$^{74}$Ge and $^{16}$O+$^{148}$Nd. The circles and solid curves denote the results of the ImQMD simulations and the entrance-channel potential with the Skyrme-energy density functional plus the ETF2 approach, respectively. The squares denote the extracted most probable barrier height from the measured fusion excitation function.  (b) and (d): empirical barrier distribution function proposed in Ref. \cite{liumin06}. }
\end{figure}

By using the ImQMD model, one can calculate the dynamical nucleus-nucleus potential \cite{ImQMD2010} in which the densities of the system and  the relative
distance $R$ between the two nuclei are functions of the evolution time.
When the projectile and target nucleus are well separated ($R\gg R_1+R_2$),
the collective relative motion plays a dominant role and the excitation energy of the reaction partners could be negligible, the nucleus-nucleus potential is thus expressed as
\begin{equation}
V_1=E_{\rm c.m.}-T_R.
\end{equation}
Where, $R_{1}$ and $R_{2}$ are the charge radii of the projectile
and the target nucleus, respectively. $T_{R}$ is
the relative motion kinetic energy of two colliding nuclei, which
can be obtained in the ImQMD simulations since the position and
momentum of each nucleon can be recorded at every time step in
the time evolutions.  After the di-nuclear system is formed
($R<R_{1}+R_{2}$), the nucleus-nucleus potential is described
by a way like the entrance channel potential \cite{Deni02}
\begin{equation}
 V_2=E_{\rm tot}(R)-\bar E_{1}-\bar E_{2},
\end{equation}
where $E_{\rm tot}(R)$ is the total intrinsic energy of the composite system which is
strongly dependent on the dynamical density distribution of the
system obtained with the ImQMD model, $\bar E_{1}$ and $\bar
E_{2}$ are the time average of the energies of the projectile and
target nuclei, respectively. Here, the values of $\bar E_{1}$ and
$\bar E_{2}$ are obtained from the energies of the projectile
(like) and target (like) nuclei in the region $R_{T}<R<R_{T}+8$.
$R_T=R_{1}+R_{2}$ denotes the touching point.  In
the calculations of $E_{\rm tot}(R)$, $\bar E_{1}$ and $\bar E_{2}$ in Eq.(8),
the extended Thomas-Fermi approximation for the intrinsic kinetic energy of the reaction system is adopted (see Refs.\cite{ImQMD2010,ImQMD2012} for details).

In this work, we write the nucleus-nucleus potential as a smooth
function between $V_1$ and $V_2$,
\begin{equation}
V_{b}(R)=\frac{1}{2}{\rm erfc}(s)V_{2}+ [1-\frac{1}{2}{\rm
erfc}(s)]V_{1}
\end{equation}
and
\begin{equation}
s=\frac{R-R_{T}+\delta}{\Delta R}
\end{equation}
with $\delta=1$ fm, $\Delta R=2$ fm. The obtained nucleus-nucleus
potential $V_{b}(R)$ approaches to $V_1$ with the increase of $R$, and
approaches to $V_2$ with the decrease of $R$.

\begin{figure}
\includegraphics[angle=0,width=0.7 \textwidth]{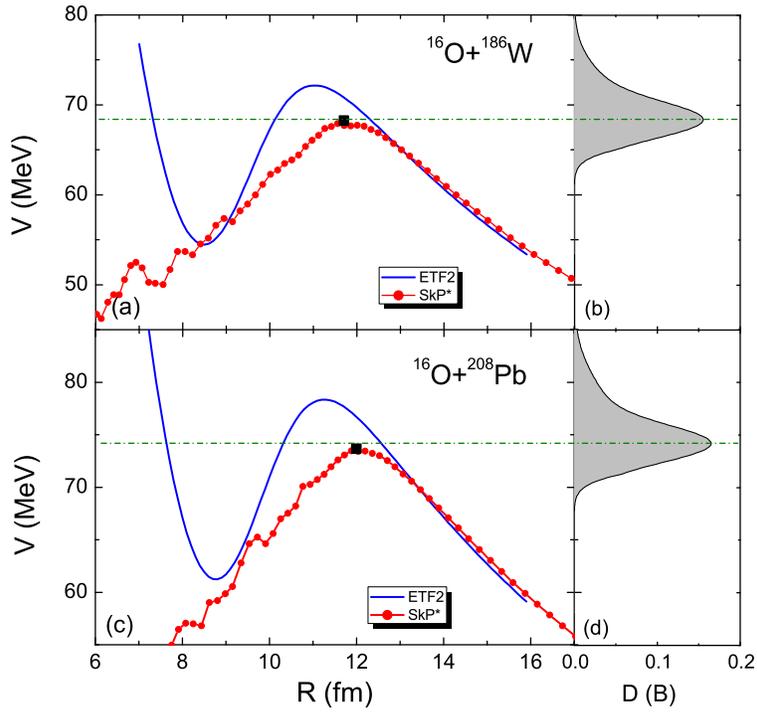}
\caption{(Color online) The same as Fig. 3, but for $^{16}$O+$^{186}$W and $^{16}$O+$^{208}$Pb.}
\end{figure}

\begin{figure}
\includegraphics[angle=0,width=0.6 \textwidth]{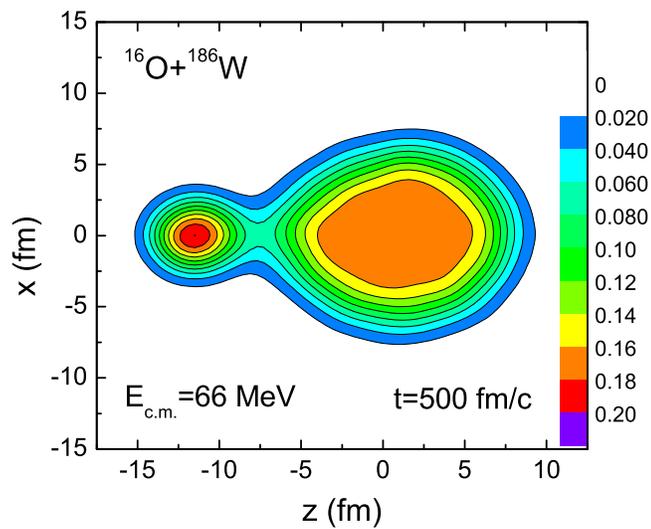}
\caption{(Color online) Density distribution of $^{16}$O+$^{186}$W at $E_{\rm c.m.}=66$ MeV and $t=500$ fm/c.}
\end{figure}

Figure 3 shows the calculated dynamical nucleus-nucleus potentials for fusion reactions $^{16}$O+$^{74}$Ge and $^{16}$O+$^{148}$Nd by using the ImQMD model with the parameter set SkP*. The blue curves denote the corresponding entrance-channel potential with the Skyrme energy-density functional plus the ETF2 approach in which the sudden approximation for the densities is used. The empirical barrier distribution functions for these two reactions are presented in Fig. 3(b) and (d). The dashed lines give the positions of the most probable barrier heights $B_{\rm m.p.}$. The black squares denote the extracted most probable barrier heights from the measured barrier distributions $D(E) =d^2 (E \sigma_{\rm fus})/dE^2$ based on the fusion excitation functions. For $^{16}$O+$^{148}$Nd, the measured data for the fusion cross sections are not many enough to extract the most probable barrier height.
It is found the dynamical barrier height from the microscopic dynamics transport model is dependent on the incident energy in the fusion reactions \cite{ImQMD2010}. Here, we set the incident energy $E_{\rm c.m.}=1.1 B_{\rm m.p.}$ in the ImQMD simulations. We found the obtained dynamical barrier height $B_{\rm dyn}\approx B_{\rm m.p.} $ at this incident energy for the fusion events. Fig. 4 show the results for  $^{16}$O+$^{186}$W and $^{16}$O+$^{208}$Pb.
From Fig. 3(a) and Fig. 4, one sees that both the dynamical barrier height at $E_{\rm c.m.}=1.1 B_{\rm m.p.}$ from the ImQMD model and the most probable barrier height from the empirical barrier distribution are close to the corresponding extracted barrier height. The static potential barriers from the sudden approximation for the densities are evidently higher but relatively thinner than the dynamical ones. To reasonably reproduce the fusion excitation functions, the empirical barrier distributions [see  the sub-figures (b) and (d) in Fig.3 and Fig.4] are proposed to take into account the nuclear structure effects and the multi-dimensional character of the realistic barrier in the ETF2 approach, and the value of the peak is lower than the corresponding barrier height from the entrance-channel potential which is based on the spherical symmetric Fermi functions for the densities of the two nuclei and the frozen-density approximation.

To understand the reduction of the potential barrier due to the dynamical effects, we show in Fig. 5 the density distribution of the fusion events in $^{16}$O+$^{186}$W at $E_{\rm c.m.}=66$ MeV which is slightly lower than the most probable barrier height of this reaction. At $t=500$ fm/c, the reaction partners locate at around the top of the dynamical barrier (the dynamical barrier height for the fusion events is about 63 MeV at $E_{\rm c.m.}=66$ MeV with SkP*). We note that the dynamical deformations of the reaction partners are evident. Comparing with the static entrance channel potential, the nose-to-nose configuration in the dynamical fusion process significantly reduces the potential barrier felt by the reaction partners.

\begin{center}
\textbf{C. Influence of Fermi constraint on fusion}
\end{center}

\begin{figure}
\includegraphics[angle=0,width=1 \textwidth]{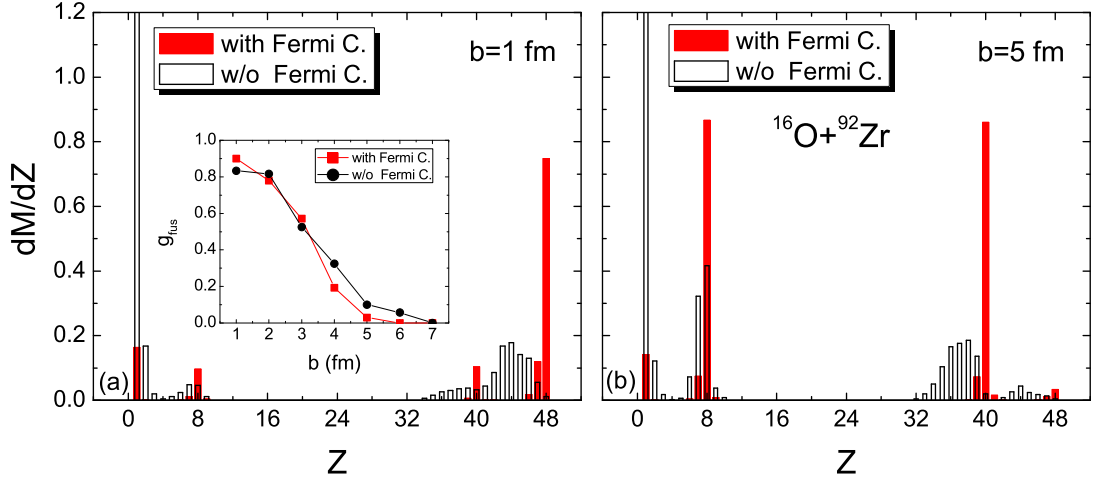}
\caption{(Color online) Charge distribution of fragments in fusion reaction $^{16}$O+$^{92}$Zr at an incident energy of $E_{\rm c.m.}=45$ MeV and $t=700$ fm/c. The filled red bars and the black hollow bars denote the results with and without the Fermi constraint being taken into account, respectively. The sub-figure in (a): fusion probability as a function of impact parameter for this reaction. }
\end{figure}

In this section, we investigate the influence of the Fermi constraint on the fusion cross sections and the fragment yields in the fusion process. As an example, we study the fusion reaction $^{16}$O+$^{92}$Zr at an incident energy of $E_{\rm c.m.}=45$ MeV which is slightly higher than the Coulomb barrier. During the propagation of nucleons, we switch off the Fermi constraint procedure in the ImQMD model at $t>300$ fm/c (the time that the two nuclei begin to touch with each other) and check the fragment yields at $t=700$ fm/c. Here, the Fermi constraint at $t\le 300$ fm/c is still considered as the same as that in the standard ImQMD simulations to guarantee the same initialization of nuclei adopted. Fig. 6 shows the charge distribution of all fragments at the central collisions with the impact parameter $b=1$ fm and the peripheral collisions with $b=5$ fm. To calculate the charge distribution, we create 500 simulation events at each impact parameter. The filled red bars and the black hollow bars denote the results with and without the Fermi constraint being taken into account, respectively. The sub-figure in Fig. 6(a) shows the fusion probability as a function of impact parameter for this reaction. When the Fermi constraint procedure is switched off, the "virtual" particle emission becomes serious and the surface diffuseness of nuclei increases. It results in that the fusion cross section increases by about 60 mb.

In addition, one can see from Fig. 6(a) that the charge distribution of the compound nucleus is wide due to the "virtual" particle emission and the peak of heavy fragments locates at about $Z=44$ rather than $Z=48$ when the Fermi constraint procedure is switched off at $t>300$ fm/c. When the Fermi constraint is taken into account during the whole time evolution, the stability of the fragments can be significantly improved and the number of "virtual" particle emission is sharply reduced. For the peripheral collisions, the scattering events are dominant from the ImQMD simulations with the Fermi constraint being considered (the fusion probability is smaller than 0.03 at $b=5$ fm, see the subfigure), and the charge numbers are $Z_p \simeq 8$ and $Z_t \simeq 40$ for the projectile (like) and target (like) fragments, respectively. It implies that the Fermi constraint is of importance for the reliable description of the fragments in heavy-ion collisions, not only for the fusion reactions, but also for the reactions at intermediate energies.

\begin{figure}
\includegraphics[angle=0,width=1 \textwidth]{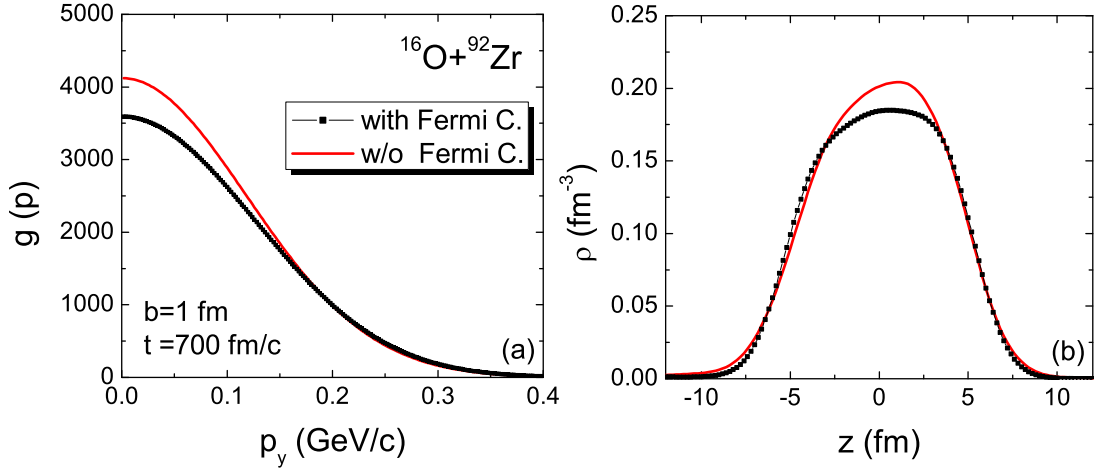}
\caption{(Color online) (a) Momentum distribution of nucleons along $p_y$-axis in $^{16}$O+$^{92}$Zr at $E_{\rm c.m.}=45$ MeV and $t=700$ fm/c. (b) Corresponding density distribution of the system along z-axis. }
\end{figure}

To understand the influence of the Fermi constraint on the reaction yields, in Fig. 7 we show the momentum and density distribution of the reaction system $^{16}$O+$^{92}$Zr at $E_{\rm c.m.}=45$ MeV, $b=1$ fm and $t=700$ fm/c. The solid curves denote the results for the cases when the Fermi constraint procedure is switched off at $t>300$ fm/c. Similar to the density distribution in the coordinate space, the momentum distribution of nucleons in the ImQMD model is expressed as
\begin{equation}
g(\mathbf{p})=\sum_i{\frac{1}{(2\pi \sigma_p^2)^{3/2}}\exp
\left [-\frac{(\mathbf{p}-\mathbf{p}_i)^2}{2\sigma_p^2} \right ]},
\end{equation}
where $\sigma_p=\frac{\hbar}{2 \sigma_r}$ represents the width of the wave packet in the momentum space.
The central values of both momentum and density distributions for the cases without the Fermi constraint being considered are much higher than those for the other cases. In the standard QMD model, the time evolution by classical equations of motion surely breaks the initial distribution which evolves into a classical Boltzmann one, even if the initial state is in agreement with the phase-space distribution of a fermionic system \cite{constrain}. From Fig. 7(a), one sees that the Fermi constraint affects the low momentum part of the momentum
distribution strongly, and can effectively restrain the number of particles with low momentum from being too large, which was also observed in \cite{ImQMD2002}. In addition to the momentum distribution, the Fermi constraint can also affect the density distribution of the reaction system. The central density is obviously higher than the normal density and the nuclear surface diffuseness is relatively larger if without the Fermi constraint. The Fermi constraint improves the momentum and density distributions of nuclear system and thus improves the stability of fragments in the ImQMD simulations.

\begin{center}
\textbf{IV. SUMMARY}
\end{center}

In summary, the heavy-ion fusion reactions induced by $^{16}$O have been systematically investigated by using the improved quantum molecular dynamics model with the parameter sets SkP* and IQ3a. The fusion cross sections at energies near and above the Coulomb barriers can be reasonably well reproduced by using this semi-classical microscopic dynamical transport model. The dynamical nucleus-nucleus potential are also simultaneously studied. The heights of dynamical fusion barriers calculated with SkP* at the incident energy of $E_{\rm c.m.}=1.1 B_{\rm m.p.}$ are close to the extracted most probable barrier height from the measured fusion excitation functions. The influence of the Fermi constraint on the fusion reactions is also investigated. The Fermi constraint plays a key role to improve the stability of the fragments and suppress the number of "virtual" particle emission, which is of importance for the reliable description of the fragments in heavy-ion collisions in addition to the mean field, not only for the heavy-ion fusion reactions at energies around the Coulomb barrier, but also for the reactions at intermediate energies. The Fermi constraint procedure can effectively suppress the central values of the momentum and density distributions of nuclear system, which helps to improve the stability of fragments in the semi-classical quantum molecular dynamics simulations. Without any adjustable model parameters and/or additional assumptions for the reaction process,  the self-consistent ImQMD model is helpful to investigate the dynamical mechanism of the reactions microscopically. For a better description of the sub-barrier fusion and the fusion reactions with heavy doubly-magic nuclei, the ImQMD model should be further improved.

\begin{center}
\textbf{ACKNOWLEDGEMENTS}
\end{center}
This work was supported by National Natural Science Foundation of
China, Nos 11275052, 11365005 and 11422548.

\end{document}